# Toward Democratizing Access to Facilities Data: A Framework for Intelligent Data Discovery and Delivery

**Yubo Qin**
Rutgers Discovery Informatics Institute, Rutgers University

**Ivan Rodero**
SCI Institute, University of Utah

**Manish Parashar**
SCI Institute, University of Utah

*Abstract*—Data collected by large-scale instruments, observatories, and sensor networks are key enablers of scientific discoveries in many disciplines. However, ensuring that these data can be accessed, integrated, and analyzed in a democratized and timely manner remains a challenge. In this article, we explore how state-of-the-art techniques for data discovery and access can be adapted to facility data and develop a conceptual framework for intelligent data access and discovery.

■ **Science** in the 21st century is being transformed by our unprecedented ability to collect and process data from a variety of sources. For example, large-scale multiuser scientific observatories, instruments, and experimental platforms provide a broad community of researchers and educators with open access to shared-use infrastructure and data products generated from geo-distributed instruments and equipment [1]. These large facilities (LF) have recently enabled significant scientific discoveries such as the detection of gravitational waves [2] and the imaging of the event horizon of a black hole [3].

However, as the number and scale of such LF increases along with corresponding growth in the number, distribution, and diversity of users, ensuring that LF data can be discovered, accessed, integrated, and analyzed in a timely manner is a growing challenge that is resulting in significant demands on LF cyberinfrastructure (CI) [4]. For example, the Ocean Observatory Initiative (OOI) [5] integrates over 1,250 instruments, producing over 25,000 data items and over 100,000 data products. Similarly, each antenna of the Square Kilometre Array (SKA), the world's largest radio telescope project, produces raw data at the rate of approximately 0.5-1TB per second and approximately 300PB of data after pre-processing per telescope per year[1].

---

[1] https://www.skatelescope.org/the-skaproject/

Satisfying the overarching goal of LF of ensuring democratized and equitable access to their data and data products across the broadest set of users can be challenging. Many LF (e.g., OOI) provide data-download portals and interfaces, and discovering data/data-product using these portals can be challenging, especially for users from a different domain, due to data complexity, diversity, and volumes. Furthermore, data access times for the same high-resolution data can range from near-real-time streaming access to several weeks via shipped disk drives. For example, Dart et al. [6] demonstrated that transferring 56TB of climate data to the NERSC computing center took up to three months due to network bandwidth and the poor performance of data transfer nodes. Furthermore, access to low-latency, high-bandwidth network connections and adequate computing and storage resources remains a significant challenge, especially for smaller and under-resourced institutions. Although national resources may be leveraged, such as those funded by the U.S. National Science Foundation (NSF), they are oversubscribed and are separate from the LF. Furthermore, using them to process LF data requires users to download the data, get the data ready for their workflows, and then upload the data and the workflow to the national resource for execution. As a result, their effective use in processing LF data is limited by users' local resources.

Consequently, democratizing LF-enabled science requires new approaches for data discovery, access, and processing. Recent years have seen advances in related technologies and capabilities aimed at increasing access to commercial data and data services. These technologies aid data discovery, proactively recommend data that are most relevant to the user, and provide anytime/anywhere access to these data. Recent efforts also address how corresponding services can be developed for science data and to support science workflows [7].

The objective of this paper is to explore how these approaches, coupled with an understanding of the data and their usage, can be effectively used to democratize access to LF data/products and to accelerate the science enabled by LF. In this paper, we build on concepts and technology presented in [9] and [10] to construct an intelligent data discovery and delivery framework composed of (1) user query analysis techniques to model access patterns and associated localities and affinities; (2) optimized data caching, data pre-fetching, and data steaming mechanisms to support optimized push-based data delivery; and (3) a data recommendation framework based on the collaborative knowledge-aware graph-attention network (CKAT) recommendation model to facilitate data discovery. We also present an evaluation[2] of the effectiveness and performance of these components using access traces from two NSF-funded large-scale observatories, the OOI and the Geodetic Facility for the Advancement of Geoscience (GAGE). The results show how the data discovery and delivery framework and the mechanisms it provides can broaden access to LF.

## AN INTELLIGENT DATA DISCOVERY AND DELIVERY FRAMEWORK FOR LF

Several research advances in CI technologies can be leveraged to address LF data discovery and delivery challenges. For example, the demilitarized zone (DMZ) network model [8] creates a dedicated network to enable high-throughput data transfer for scientific data flows. A data transfer node (DTN) [9] provides an access point for users connecting to a DMZ network and is responsible for managing and optimizing data transfers. These CI elements can be used to analyze users' requests in the network; identify the patterns, localities, and affinities described above; and host services that use this information to improve data access performance. For example, frequently accessed data can be cached at DTNs. Furthermore, the analysis can be used to develop strategies for predicting future queries and for pre-fetching data to DTN storage closer to the user. Finally, the analysis of user data query patterns can be used for recommending other relevant data to users and to host such recommendation services at DTNs. Motivated by these observations, we propose an intelligent data discovery and delivery framework as illustrated in **Figure 1**. The *Analysis Module* analyzes user accesses to identify patterns as well as localities and affinities. This knowledge is then used by the *Intelligent Data Delivery Service* to cache frequently used data, predict a user's subsequent queries, and pre-fetch the corresponding data. The *Data Discovery Service* uses the analysis along with domain-specific data models and user associations to generate a knowledge graph to implement a data recommendation service. Together, these services can help move us toward democratizing access to LF data following the FAIR data access principles.

---

[2] The sources are available at https://gitlab.com/sci-data for reproducibility.

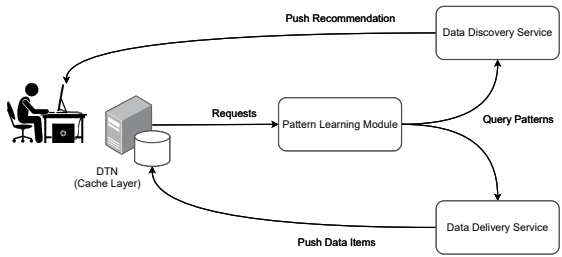

**Figure 1. Overarching architecture of the intelligent data service framework.**

The implementation of this framework leverages the Virtual Data Collaboratory (VDC). VDC implements a data-DMZ that supports data sharing and data-driven collaborations between VDC hubs through DTNs. It also integrates data from other sources such as LFs.

ANALYZING LF DATA AND DATA ACCESS

Analyzing LF data and user data usage and query patterns is essential to understanding correlations and predicting user behaviors. Such analysis also enables the identification of ineffective practices or bottlenecks and thus the improvement of system performance. Our goal is to classify users and model their queries or requests to identify affinities that can anticipate users' requests, thus enabling us to discover relevant data items. To do this, we analyzed one year of requests from the OOI and GAGE access traces. OOI and GAGE support data discovery and access through their web data portal and API. Our analysis showed that wheras most of the accesses in the traces were via the data portal, 90.1% of the data downloads used APIs and were triggered by workflows or scripts. We term users accessing data using these APIs *program users*, and since this type of user is the major data consumer, we focus on improving the access performance for program users. We believe that any request patterns identified for these requests would remain consistent since they are programming based on well-defined requirements, and they can be used to predict future requests based on historical data.

We analyzed the program requests in the traces based on different parameters, such as time intervals between them, set of queried data, and query time range to identify consistent request routines. We identified three access patterns: regular, overlapping, and real-time requests. Regular requests represent the most common request type and query new data since the last request without any overlap. Real-time requests are regular high-frequency requests typically used to monitor the occurrence of specific events. Overlapping requests are like regular requests but with overlaps across consecutive requests.

We also found significant data reuse across queries in both traces. On the one hand, this reuse comes from the overlapping portion of a user's consecutive requests. On the other hand, it results from similar data requests generated by different users (i.e., groups of users request similar data items). This reuse allows us to leverage data-caching mechanisms immediately to improve data access performance and reduce redundant queries and transfers.

We also analyzed correlations across data queries and identified three key types of affinities:

1. *Facility instrument locality*. Data from instruments that are located close together tend to be queried together. LFs typically deploy multiple instruments in an area with high research value. As a result, users studying that area will naturally download data from some or all the instruments within the area. Consequently, instrument locality defines spatial affinities between data and data products and results in corresponding correlations across data queries. Our analysis of the OOI and GAGE access traces shows that, on average, users make 43.1% and 36.3% of their queries for data objects from instruments located in one region, and 51.6% and 68.8% of their queries are to the same data type, respectively. We also observed a temporal correlation across user queries in our analysis. **Figure 2** shows the requested data objects (using their instrument name and the instrument location) from a fragment of the OOI trace clustered by users. The observed patterns suggest the existence of a spatial correlation across the requests as users request multiple data objects of one region and the same type of data object in nearby regions. We also observed a temporal correlation in our analysis.

2. *Domain data model*. Data produced by LF instruments and observations are typically used to derive data products, and the "recipes" used in this derivation are defined in the facilities' data models. For example, studies in oceanology use conductivity, temperature, and depth data to calculate water salinity and density. These domain-specific relationships define data-model affinities between data and result in corresponding correlations across data queries. For example, conductivity, temperature, and depth data are correlated and are likely to be requested together to calculate water salinity and density.

3. *User association*. A classic association used in collaborative filtering recommendation models is that users with similar interests download similar data items. This association indicates that if two users have similar characteristics, such as research interests, they will likely request the same data items. Identifying such associations is difficult since facilities typically do not keep track of users or ask them to create profiles. However, our analysis shows that determining user similarity according to their geographical proximity is possible because most LF users are researchers or other stakeholders who are part of larger research groups and/or working as part of projects. Thus, users working on the same project (e.g., from the same organization or institute) would, with high probability, request similar data items. Consequently, we can leverage user locality as an indicator of data affinity and the resulting correlation across queries. Our analysis of the OOI and GAGE access traces validates user association and shows that users within the same research group (or same organization) tend to have similar data-query patterns.

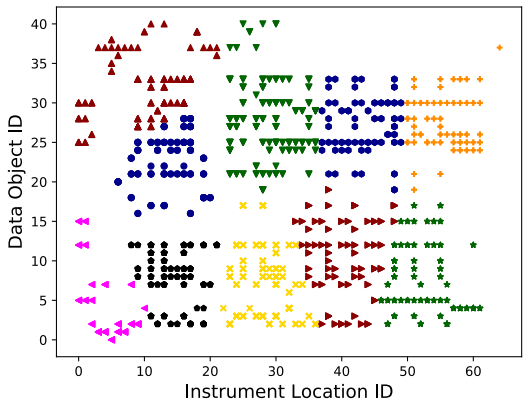

**Figure 2. Representation of users' requests from a fragment of the OOI trace. Each cluster represents a distinct user (please see in color).**

INTELLIGENT DATA DELIVERY SERVICE

The data delivery service aims to improve the user data access performance by pre-fetching data to DTNs close to the users and enabling them to access data primarily from the cache rather than retrieving them from the remote data source. The pre-fetching mechanism is based on user request history. As discussed above, over 90.1% of the volume of data downloaded is in response to queries from program users, i.e., queries generated by automated programs or scripts. These queries, by their nature, are predictable. As a result, by pre-fetching the relevant data items and caching requested data that can potentially service future requests, the local cache at the DTN storage can serve a large fraction of user requests.

The data delivery service is designed based on these insights, as illustrated in **Figure 3**. The architecture consists of two primary functional components: the cache layer and the data push mechanism. The cache layer spans DTNs at the data sources (i.e., the LFs) and at the users, and forms a distributed interconnected cache network using storage available at the DTNs. The goal of the data placement strategy is to place the data at local DTNs that are close to potential users, and to keep data with high probability to be accessed in the future in the cache network. The overall data placement strategy is composed of local caching based on LRU and virtual groups. Virtual groups are groups of users who have common data interests and are geographically close to each other. We can place data objects of interest to a virtual group at a DTN that has the best network connected to the corresponding set of users. We use k-means clustering to identify virtual groups. The data push mechanism is responsible for pre-fetching and stream data based on the analysis user access patterns.

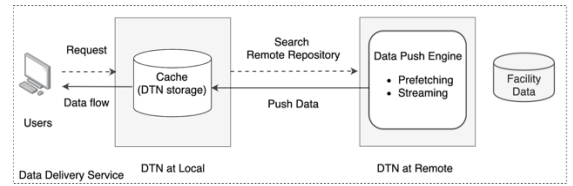

**Figure 3. Architecture of the intelligent data delivery service.**

Clients run on DTNs at the user side and pre-process user requests by searching for the requested data in the cache layer. If the requested data are not present in the cache, the request is forwarded to a server, which runs at DTNs at the data source, i.e., the LF. The server also runs the data pre-fetching mechanisms and manages the placement of cached data.

The evaluation of the data delivery service is based on the simulation of the VDC architecture with seven geographically distributed DTNs interconnected via a wide-area network and uses the OOI and GAGE request traces to evaluate the effectiveness and performance of the data delivery service under various operational conditions and using different scenarios. The results show that the delivery service improves data delivery performance and quality of service along different dimensions as compared to current practices. Key results of our evaluation are summarized below.

1. *Higher throughput and lower latencies achieved*. The data delivery service provides improvements in the data delivery throughput of over 2,600 times and reduces the latency from request submission to data access by 38%. These improvements are mainly the result of requests being served using cached data.
2. *Data delivery performance is more robust to network variation*. The data delivery service is more robust to the network variations. Data access performance does not dramatically change with changing network conditions. Overall network bandwidth requirements are also reduced, especially over the wide area, as redundant data requests are eliminated.
3. *Reduced load and network traffic at the LF*. The data delivery service reduces the total requests and corresponding network traffic at the LF, as many requests are satisfied using cached data at the client side.

We also studied how pre-fetching improves local data reuse. **Figure 4** shows the average percentage of requests that are served from a local cache, using virtual groups, and caching with pre-fetching (referred to as "Smart Cache" in **Figure 4)**. The results indicate that pre-fetching enables users to obtain more data from their local cache. As opposed to passively searching cached data, the pre-fetching mechanism proactively pushes data toward to user. Thus, it ensures that users can access more of their data locally regardless of whether they are reused from the previous requests. Furthermore, the pre-fetching mechanism can achieve near-optimal performance with a small cache size. Please refer to [10] for more details.

Overall, we observe that the proposed data delivery service significantly improves performance and enhances its service robustness in response to complex real-world variations.

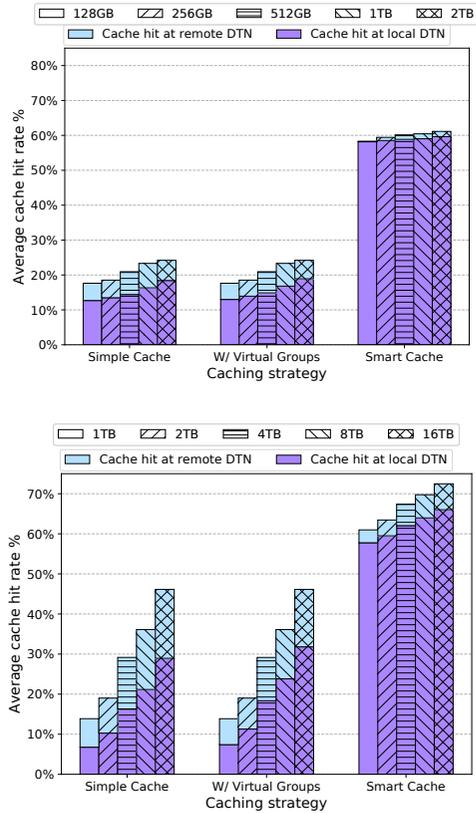

**Figure 4. Percentage of data movement from the local cache for OOI (top) and GAGE (bottom).**

## INTELLIGENT DATA DISCOVERY

As noted earlier, as the number of data and data-products available at LF grows, discovering data/data-products of interest can be extremely challenging. The data discovery service aims to recommend data and data products to users that are most relevant to their research interests. However, most popular e-commerce recommendation models are based on linked data and rich metadata about a user's personal history and preferences. Such data may not be available and relevant when recommending data and data products from an LF, and the existing models do not directly translate for such recommendations. In the case of LF users, data requests are based on research needs. Furthermore, facilities typically do not keep track of user histories or require users to create personal profiles listing their preferences. As a result, the data discovery service uses knowledge about user-query patterns and correlations across user queries along with domain-specific data models.

As noted earlier, our analysis of user requests to production facilities identified three key affinities that characterize query behaviors: instrument locality, data-domain model, and user association. Harvesting these affinities is critical to automate the data discovery process, and they can be obtained by a combination of information sources, including the facility instrument metadata, user query traces, and external sources such as Wikipedia. This information is then captured in a knowledge graph, which is an effective method to represent such information, capturing the facts as the nodes and presenting the relationship among facts as the paths. Several recently proposed recommendation models leverage knowledge graphs to carry the auxiliary information to help address the cold-start and data-sparsity challenges. In our case, the knowledge graph contains information about the three types of affinities described above.

To construct data discovery services capable of recommending relevant data items to LF users, we developed a Collaborative Knowledge-aware ATtention network (CKAT) recommendation model. The overall recommendation generation process is summarized in **Figure 5**.

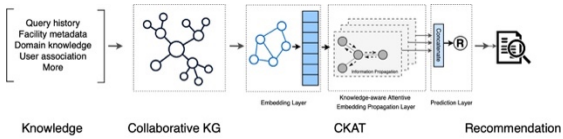

**Figure 5. Overview of the recommendation process based on based on the CKAT model.**

We first represent each knowledge source as its own graph. To correlate the knowledge sources, we can consolidate these individual knowledge graphs into a Collaborative Knowledge Graph (CKG) through entity alignment. The CKG enables diverse information to be connected in the graph to form a collaborative signal. The CKG construction process also allows us to examine different knowledge combinations, which is key to achieving sound recommendations. Paths in the CKG represent the connection of two data items. Whereas first-order connectivity occurs when data items are directly connected, high-order connectivity occurs when there is a path between two data items across multiple nodes in the graph. The advantage of combining various knowledge sources using the CKG is the ability to identify connections between two indirectly related data items, which requires capturing long-distance paths (i.e., high-order connectivity) in the graph, which can be achieved using a graph neural network (GNN). However, before sending the CKG to the GNN, we need to convert and embed the graph representation into a vector representation, as illustrated in **Figure 6**.

Existing embedding models such as TransE and TransH assume that entity and relation are vectors in the same space, so similar entities will be close to each other in the same entity space. However, each entity can have many aspects, and different relations pay attention to the various aspects of the entity. For example, the relation of (location, contains, location) is 'contains', and the relation of (person, born, date) is 'born'. Therefore, these two relations are very different. To address this issue, we select the TransR [11] model, which considers relations in two distinct spaces, i.e., entity space and multiple relation spaces (relation-specific entity spaces), and performs the translation in the corresponding relation space, thus reflecting the importance of two items in different relationships.

In the case of LF data usage, two data items can be used together for different research purposes for which the correlation or importance of the two data items is different. For example, the correlation of physical environmental variables can be relevant for climate change research but can also help in the understanding of shorter term effects such as the impact of invasive species on migratory species through predation. Therefore, being able to distinguish between these differences is important. Recent work by Wang et al. [12] has demonstrated that GNNs can capture high-order connectivity through high-order information propagation. However, key issues may impact learning performance, such as irrelevant paths (i.e., noise) that can impact the ability to find actual correlations. Since nodes can be connected via different paths in the graph, not all of them have the same importance in a certain relation. Thus, we apply the attention mechanism to enable the GNN training process to focus on the important relations.

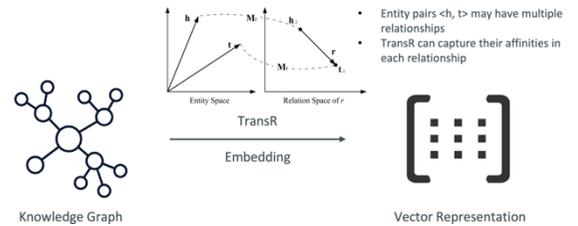

**Figure 6. Overview of the embedding layer (i.e., converting a KG into vector representation)**

The experimental evaluation of the CKAT-based data discovery service shows that it can effectively recommend data to users, and it has recommendation accuracy that is over 6.12% higher compared to the state-of-the-art models. The evaluation also shows that knowledge combination plays a key role in the results, indicating knowledge sources must be carefully selected to obtain sound results. The results also differed between facilities, indicating that a pre-selection process is needed for each facility to achieve optimal results. More knowledge sources do not necessarily provide better recommendation results. Only the related knowledge sources are needed. Our experiments illustrate that once we purposely insert "noise" (i.e., irrelevant) knowledge to the best knowledge combination, the recommendation worsens, which emphasizes how important the knowledge selection process is. Furthermore, when we disable the attention mechanism, the recommendation result is impacted by every knowledge source input, which illustrates the attention mechanism does help eliminate the noise knowledge information and helps improve training accuracy. **Figure 7** illustrates the high-order connectivity in inferring user preferences, i.e., using the attention score to represent the affinity using OOI data. The figure shows how CTD data are recommended when the user previously queried ADCP data, which are obtained from sensors at the same location (Cabled Endurance Array). We observed that the instrument locality is more influential than other general attributes.

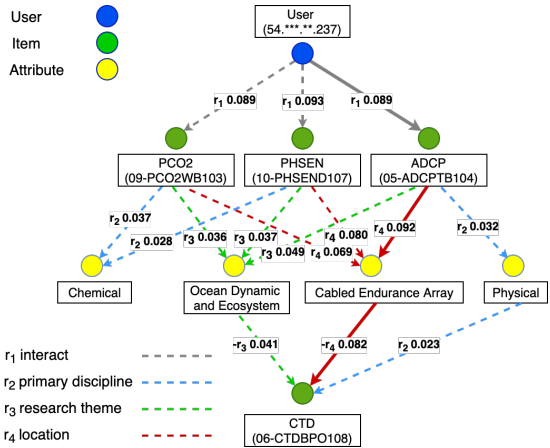

**Figure 7. Sample recommendation outcome using OOI data. The arrows show the recommendation scores.**

The CKAT model demonstrates a new methodology and direction to assist users in discovering facility data through exploiting diverse knowledge sources. Please refer to [10] for more details.

## CONCLUSION

Society today is facing more complex problems requiring the solutions and associated data-driven workflows to discover sufficient correlated data as the inputs, which often involve multiple data sources across disciplines. LF data and CI are becoming an essential part of data-driven science and will play more important roles in future scientific discoveries. We propose an approach to address such data discovery and access challenges in using the facility data by exploring the knowledge of user access behaviors and the opportunities in the CI to bring convenience to the science communities and democratize access to LFs data and knowledge.

This article provides a conceptual data framework for intelligent data access and delivery. It proposes the concept and construction methodology of the collaborative knowledge graph to represent the auxiliary information extracted from facility metadata, domain knowledge, and history of user requests. The experimental evaluation building upon the VDC architecture and modern AI-based models exhibits a significant reduction of data access and data transfers from LFs' repositories compared to current approaches.

The path forward to improve data access and delivery methods includes distilling knowledge exploring the connection between LFs data and their associated research, which is natural in human learning. In addition, these methods can deliver personalized recommendations by creating researchers' profiles from publications and modern techniques such as natural language processing and knowledge representation learning.

These concepts can translate to other LFs beyond those studied in this article. We envision the proposed framework through VDC to become a pervasive in-network environment that can enable all service components with online learning capabilities, from a distributed cache network to hybrid pre-fetching models, data streaming mechanisms, and intelligent cache data placement strategies. VDC represents the foundation to enable the proposed framework and recommendation-based system across multiple LFs as a central component of a national data fabric.

## ACKNOWLEDGMENT

This research is supported in part by NSF via grants numbers OAC 1835692 and OAC 1640834.

# REFERENCES

1. I. Rodero and M. Parashar, "Data Cyberinfrastructure for End-to-end Science," *Computing in Science & Engineering*, vol. 22, no. 5, pp. 60–71, 2020.
2. B. P. Abbott, et al., "Observation of gravitational waves from a binary black hole merger," *Physical review letters*, vol. 116, no. 6, p. 061102, 2016.
3. K. Akiyama, et al., "First m87 event horizon telescope results. IV. Imaging the central supermassive black hole," *Astrophys. J. Lett*. 875 (1) (2019) L4.
4. M. Parashar, et al., "The Virtual Data Collaboratory: A Regional Cyberinfrastructure for Collaborative Data-Driven Research," *Computing in Science & Engineering*, vol. 22, no. 3, pp. 79–92, 2020.
5. M. Smith, et al., "The Ocean Observatories Initiative," *Oceanography*, vol. 31, no. 1, pp. 16–35, 2018.
7. E. Dart and M. F. Wehner, "An Assessment of Data Transfer Performance for Large-Scale Climate Data Analysis and Recommendations for the Data Infrastructure for CMIP6," *arXiv:1709.09575*, 2017.
7. K. Fauvel, et al., "A Distributed Multi-sensor Machine Learning Approach to Earthquake Early Warning," *34th AAAI Conf. on Artificial Intelligence*, pp. 403-411, 2020.
8. L. Smarr, et al. "The pacific research platform: Making highspeed networking a reality for the scientist," *Practice and Experience on Advanced Research Computing*, pp. 29:1–29:8, 2018.
9. Y. Qin, et al., "Leveraging User Access Patterns and Advanced Cyberinfrastructure to Accelerate Data Delivery from Shared-use Scientific Observatories," *Future Generation Computer Systems*, vol. 122, pp. 14–27, 2021.
10. Y. Qin, I. Rodero, and M. Parashar, "Facilitating Data Discovery for Large-scale Science Facilities using Knowledge Networks," *IEEE International Parallel and Distributed Processing Symposium*, pp. 651-660, 2021.
11. Y. Lin, et al., "Learning Entity and Relation Embeddings for Knowledge Graph Completion," *29th AAAI Conf. on Artificial Intelligence*, pp. 2181-2187, 2015.
12. X. Wang, et al., "KGAT: Knowledge Graph Attention Network for Recommendation," *25th ACM SIGKDD International Conference on Knowledge Discovery & Data Mining*, pp. 950-958, 2019.

**Yubo Qin** received his Ph.D. degree from Rutgers University. His research focused on addressing data discovery and geo-distributed data sharing challenges. Contact him at yubo.qin@rutgers.edu.

**Ivan Rodero,** is Research Computer Scientist at the Scientific Computing and Imaging (SCI) Institute at the University of Utah. His research focusses on data-driven science and engineering, high performance parallel and distributed computing, and advanced cyberinfrastructure. He is senior member of IEEE, and ACM. Contact him at ivan.rodero@utah.edu.

**Manish Parashar,** is the Director and Chair in Computational Science and Engineering at the Scientific Computing and Imaging (SCI) Institute and Professor at the School of Computing at the University of Utah. His research interests are in the broad areas of parallel and distributed computing and computational and data-enabled science and engineering. He is Fellow of the AAAS, ACM, and IEEE. Contact him at manish.parashar@utah.edu.